\documentclass[fleqn,10pt]{wlscirep}
\usepackage[utf8]{inputenc}
\usepackage[T1]{fontenc}
\usepackage{multicol}
\usepackage{graphicx}
\newenvironment{Figure}
  {\par\medskip\noindent\minipage{\linewidth}}
  {\endminipage\par\medskip}
  
\usepackage{mathtools}
\usepackage{amsmath}
\usepackage[export]{adjustbox}

\usepackage[flushleft]{threeparttable}

\usepackage{tikz}
\usepackage{textcomp}
\usepackage{lipsum}
\usepackage[normalem]{ulem} 
\usepackage{url}

\newcommand{\norm}[1]{{||{#1}||}}

\title{SyntheX: Scaling Up Learning-based X-ray Image Analysis Through In Silico Experiments}

\author[1,*]{Cong~Gao}
\author[1]{Benjamin~D.~Killeen}
\author[1]{Yicheng~Hu}
\author[1]{Robert~B.~Grupp}
\author[1]{Russell~H.~Taylor}
\author[1]{Mehran~Armand}
\author[1,*]{Mathias~Unberath}
\affil{Johns Hopkins University, Baltimore, MD, USA}
\affil[*]{Send correspondence to: \{cgao11,mathias\}@jhu.edu}

\begin{abstract}
Artificial intelligence (AI) now enables automated interpretation of medical images for clinical use. However, AI’s potential use for interventional images (versus those involved in triage or diagnosis), such as for guidance during surgery, remains largely untapped. This is because surgical AI systems are currently trained using \emph{post hoc} analysis of data collected during live surgeries, which has fundamental and practical limitations, including ethical considerations, expense, scalability, data integrity, and a lack of ground truth.

Here, we demonstrate that creating realistic simulated images from human models is a viable alternative and complement to large-scale \emph{in situ} data collection. We show that training AI image analysis models on realistically synthesized data, combined with contemporary domain generalization or adaptation techniques, results in models that on real data perform comparably to models trained on a \emph{precisely matched} real data training set. Because synthetic generation of training data from human-based models scales easily, we find that our model transfer paradigm for X-ray image analysis, which we refer to as \emph{SyntheX}, can even outperform real data-trained models due to the effectiveness of training on a larger dataset. We demonstrate the potential of \emph{SyntheX} on three clinical tasks: Hip image analysis, surgical robotic tool detection, and COVID-19 lung lesion segmentation.  

SyntheX provides an opportunity to drastically accelerate the conception, design, and evaluation of intelligent systems for X-ray-based medicine. In addition, simulated image environments provide the opportunity to test novel instrumentation, design complementary surgical approaches, and envision novel techniques that improve outcomes, save time, or mitigate human error, freed from the ethical and practical considerations of live human data collection. 
\end{abstract}

\begin{document}
\flushbottom
\maketitle
\thispagestyle{empty}

\begin{multicols}{2}

\section{Main}

\subsection{Background}

\begin{figure*}[t!]
    \centering
    \includegraphics[width=1.0\textwidth]{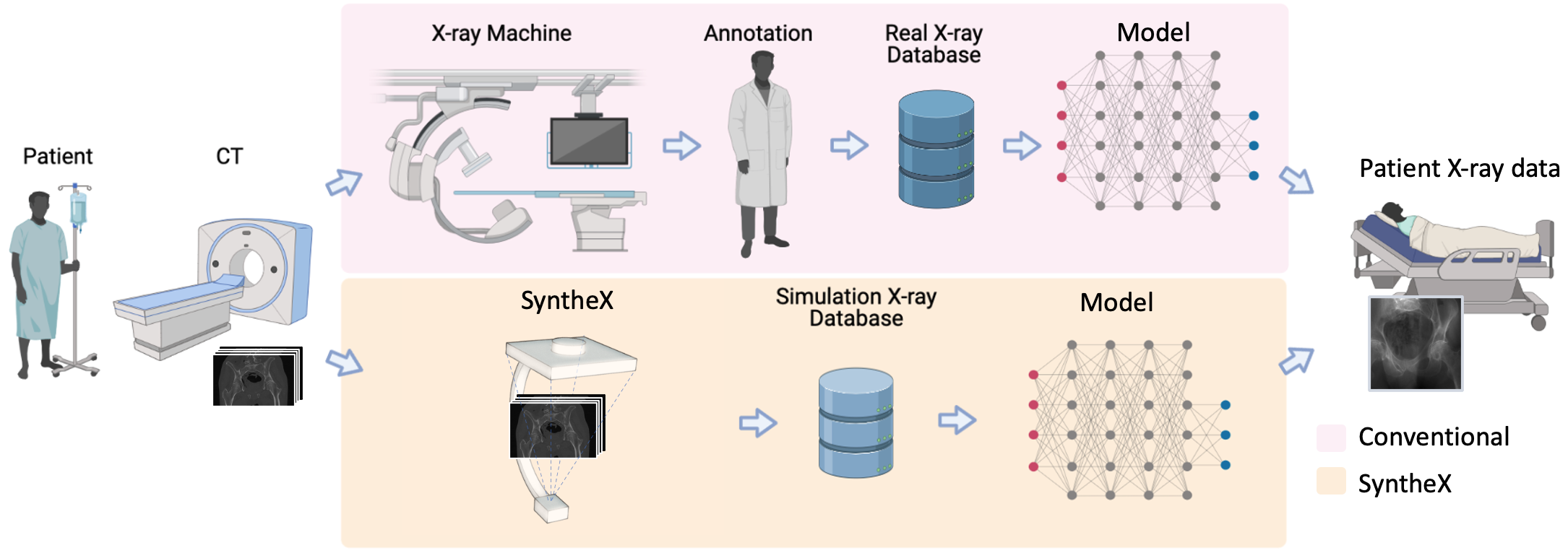}
    \caption{Overall concept of SyntheX. Pink: Conventional approach for learning-based tasks on medical imaging. Curating a relevant database of real X-ray samples requires real data acquisition and costly annotation from domain experts. Yellow: SyntheX enables simplified and scaled up data curation because data generation is synthetic and synthesized data can be annotated automatically through propagation from the 3D model. SyntheX results in deep learning image analysis models that perform comparably to or better than real-data trained models.}
    \label{fig:overview}
\end{figure*}

Advances in robotics and artificial intelligence (AI) are bringing autonomous surgical systems closer to reality. However, developing the AI backbones of such systems currently depends on collecting training data during routine surgeries. This remains one of the largest barriers to widespread use of AI systems in interventional clinical settings, versus triage or diagnostic settings, as the acquisition and annotation of interventional data is time intensive and costly. Furthermore, while this approach can contribute to the automation or streamlining of \emph{existing} surgical workflows, robotic and autonomous systems promise even more substantial advances: novel and super-human techniques that improve outcomes, save time, or mitigate human error. This is perhaps the most exciting frontier of computer-assisted intervention research. 

Conventional approaches for curating data for AI development (i.e., sourcing it retroactively from clinical practice) are insufficient for training AI models that benefit interventions that use novel instrumentation, different access points, or more flexible imaging. This is because they are, by definition, incompatible with contemporary clinical practices and such data does not emerge from routine care. Furthermore, these novel systems are not readily approved, and thus not easily or quickly introduced into clinical practice. \emph{Ex vivo} experimentation does not suffer the same ethical constraints; however, it is costly and requires mature prototypes, and therefore does not scale well.

A promising alternative to these strategies is simulation, i.e., the \emph{in silico} generation of synthetic interventional training data and imagery from human models. Simulation offers a rich environment for training both human and machine surgeons alike, and sidesteps ethical considerations that arise when exploring procedures outside the standard of care. Perhaps most importantly, \emph{in silico} surgical sandboxes enable rapid prototyping during the research phase. Simulation paradigms are inexpensive, scalable, and rich with information. While intra-operative data is generated in highly unstructured and uncontrolled environments, and requires manual annotation, simulation can provide detailed ground truth data for every element of the procedure, including tool and anatomy pose, which are invaluable for AI development.

However, simulations can fall short of real surgery in one key aspect: realism. The difference in characteristics between real and simulated data is commonly referred to as the ``domain gap.'' The ability of an AI model to perform on data from a different domain, i.\,e. with a domain gap from the data it was trained on, is called ``domain generalization.'' Domain gaps are problematic because of the well documented brittleness of AI systems~\cite{drenkow2021robustness}, which exhibit vastly deteriorated performance across domain gaps. This may happen even with simple differences, such as noise statistics, contrast level, and other minutiae~\cite{taori2020measuring, mahmood2018deep, gu2020extended, bier2019learning}. This unfortunate circumstance, which applies to all machine learning tasks, has motivated research in the AI field on \emph{Sim2Real} transfer, the development of domain transfer methods. 

In this work, we describe advances in X-ray simulation and domain transfer methods, which, when combined, contribute feasible solutions for training AI algorithms on synthetic data while preserving their performance under domain shift for evaluation and deployment in the real world. The overall concept of training deep neural networks on realistically simulated data from annotated CT scans using domain randomization, which we refer to as \emph{SyntheX}, is illustrated in Fig.~\ref{fig:overview} and we demonstrate its utility and validity on three clinical applications: hip imaging, surgical robotic tool detection and COVID-19 lesion segmentation. 

At the core of our report is an experiment on precisely controlled data that isolates and quantifies the effect of domain shift for AI-based X-ray image analysis. Using CT images from human cadavers and corresponding C-arm X-ray images acquired during surgical exploration, we generated a hip image dataset consisting of geometrically identical images across various synthetic and the real domains to train AI models for hip image analysis. To our knowledge, no study to date has isolated the effect of domain generalization using precisely matched datasets across domains. This work also, for the first time, demonstrates a feasible and cost-effective way to train AI image analysis models for clinical intervention on synthetic data in a way that provides comparable performance to training on real clinical data in multiple applications.

\begin{figure*}[b!]
    \centering
    \includegraphics[width=0.95\textwidth]{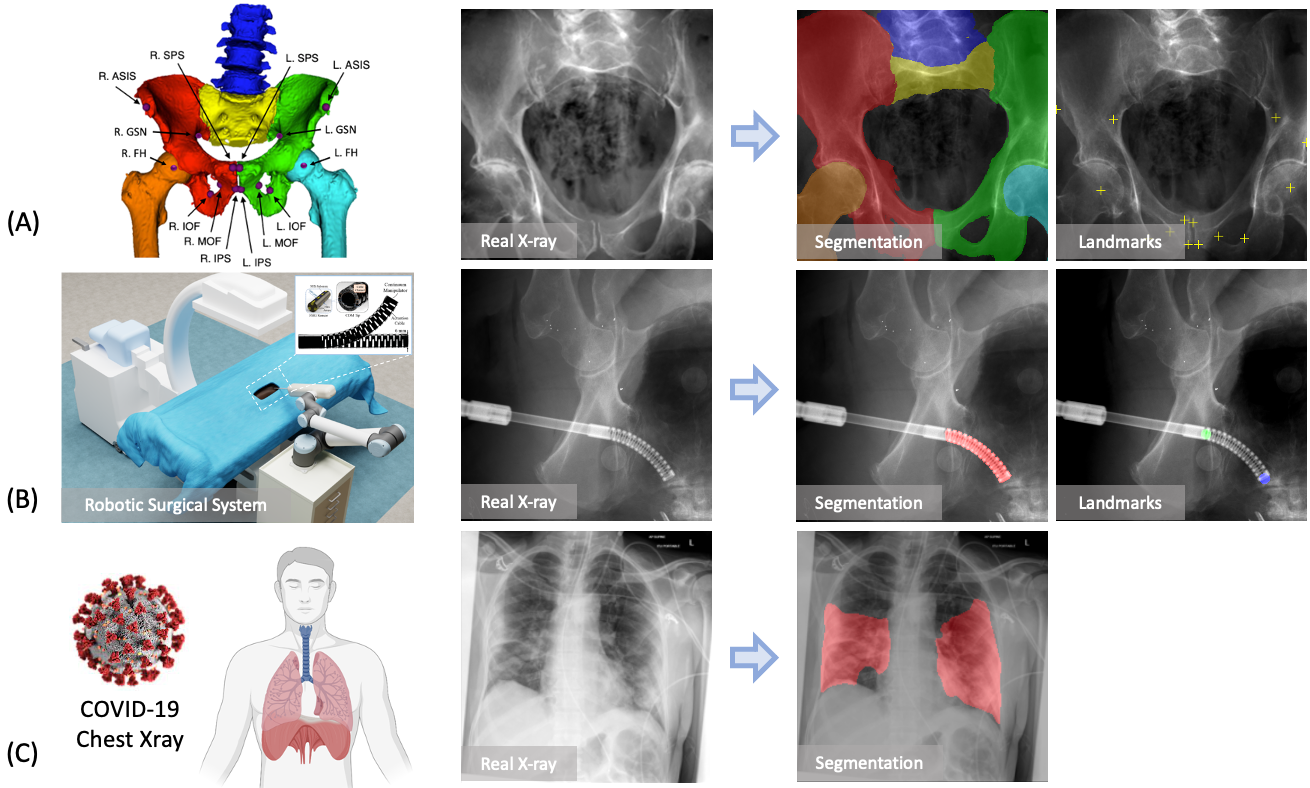}
    \caption{(A) Hip imaging. The hip anatomical structures include left and right hemipelvis, lumbar vertebrae, upper sacrum, left and right femurs. The anatomical landmarks consist of left and right anterior superior iliac spine~(ASIS), center of femoral head~(FH), superior pubic symphysis~(SPS), inferior pubic symphysis~(IPS), medial obturator foramen~(MOF), inferior obturator foramen~(IOF), and the greater sciatic notch~(GSN). These landmarks are useful in identifying the anterior pelvic plane~(APP) and initializing the 2D/3D registration of both pelvis and femur~\cite{gao2020tmrb, nikou2000description}. (B) Surgical robotic tool detection. An illustration of the image-guided robotic surgical system is shown on the left. An example real X-ray image and the corresponding segmentation and landmarks of the continuum manipulator is shown on the right. (C) COVID Chest X-ray lesion segmentation. A real COVID-19 infected chest X-ray image is presented with its lesion segmentation mask.}
    \label{fig:clinical_task}
\end{figure*}

\subsection{Clinical Tasks}
\label{sec:clinical-task}
We demonstrate the benefits of SyntheX on three X-ray image analysis downstream tasks: hip imaging, surgical robotic tool detection and COVID-19 lesion segmentation in chest X-ray~(Fig.~\ref{fig:clinical_task}). All of the three tasks use deep neural networks to make clinically meaningful predictions on X-ray images. We introduce the clinical motivations for each task in the following sections. Details of the deep network and training/evaluation paradigm are described in Section~\ref{sec:model_evaluation}.

\subsubsection{Hip Imaging}
Computer-assisted surgical systems for X-ray-based image guidance have been developed for trauma surgery~\cite{vleung2010image}, total hip arthroplasty~\cite{kelley2009role}, knee surgery~\cite{kordon2019multi}, femoroplasty~\cite{gao2020fiducial}, pelvis osteotomy~\cite{grupp2019pose}, and spine surgery~\cite{nolte2000new}. The main challenge in these procedures is to facilitate intra-operative image-based navigation by continually recovering the spatial tool-to-tissue relationships from 2D transmission X-ray images. One effective approach to achieving spatial alignment is the identification of known structures and landmarks in the 2D X-ray image, which then are used to infer poses~\cite{grupp2020automatic,unberath2021impact}.

In the context of hip imaging, we define six anatomical structures and fourteen landmarks as the most relevant known structures. They are shown in Fig.~\ref{fig:clinical_task}~(A). We trained deep networks using SyntheX to make these detections on X-ray images. Synthetic images were generated using CT scans selected from the New Mexico Decedent Image Database~\cite{newmexico}. The 3D anatomical landmarks were manually annotated and the anatomical structures were segmented using the automatic method described in~\cite{krvcah2011fully}, which were then projected to 2D as labels following the simulation X-ray geometries. We evaluate the performance of our model on 366 real X-ray images collected from six cadaveric specimens. On real images, ground truth target structures were annotated semi-automatically. This real dataset also serves as the basis for our precisely controlled experiments that isolate the effect of the domain gap. We provide substantially more details on the creation,annotation, and synthetic duplication of this dataset in Section~\ref{sec:pelvic_dataset}. 

\subsubsection{Surgical Robotic Tool Detection}
Automatic detection of the surgical tool from intra-operative images is an important step for robot-assisted surgery since it enables vision-based control~\cite{bouget2017vision}. Because training a detection model requires sufficient image data with ground truth labels, developing such models is only possible after the surgical robot is mature and deployed clinically. We demonstrate AI model development for custom and pre-clinical surgical robotic tools. 

We consider a continuum manipulator as the target object. Continuum manipulators (CMs) have been investigated in minimally-invasive robot-assisted orthopedic procedures because of their significant dexterity and stiffness~\cite{alambeigi2017curved, bakhtiarinejad2019biomechanical}, but they are not currently used clinically nor easily manufactured for extensive cadaveric testing. Using SyntheX, we address CM detection, which consists of segmenting the CM body and predicting distinct landmarks in the X-ray images. The semantic segmentation mask covers the 27 alternating notches which discern the CM from the other surgical tools; the landmarks are defined as the start and end points of the CM centerline~\cite{gao2021fluoroscopic}. Synthetic images were generated using CT scans selected from the New Mexico Decedent Image Database~\cite{newmexico} and a computer-aided design model of the CM. 3D CM segmentations and landmark locations were determined through forward kinematics and then projected to 2D as training labels using the X-ray geometry. The performance was evaluated on 264 real X-ray images of the CM during pre-clinical cadaveric testing. On real images, ground truth segmentation masks and landmark locations were annotated manually.

\subsubsection{COVID-19 Lesion Segmentation}
Chest X-ray~(CXR) has emerged as a major tool to assist in COVID-19 diagnosis and guide treatment. Numerous studies have proposed the use of AI models for COVID-19 diagnosis from CXR and efforts to collect and annotate large amounts of chest X-rays are underway. Annotating these images in 2D is expensive and fundamentally limited in its accuracy due to the integrative nature of X-ray transmission imaging. While localizing COVID-19 presence is possible, deriving quantitative chest X-ray analysis solely from chest X-rays is impossible. Given the availability of CT scans of patients suffering from COVID-19, we demonstrate lung imaging applications using SyntheX. 

Specifically, we consider the task of COVID-19 lesion segmentation which is possible also from chest X-ray to enable comparison. We used the open source COVID-19 CT dataset released by ImagEng lab~\cite{zaffino2021open} and the CT scans released by UESTC~\cite{wang2020noise} to generate synthetic CXRs. A 3D infection mask was created for each CT using the automatic lesion segmentation method COPLE-Net~\cite{wang2020noise}. Synthetic images and labels were generated using the paired CT scan and segmentation mask. The segmentation performance was tested on the benchmark dataset QaTa-COV19~\cite{degerli2021covid}, which contains 2,951 real COVID-19 CXR samples. Ground truth segmentation masks for the COVID-19 lesions in these CXRs are supplied with the benchmark, and were created in a human–machine collaborative approach.



\begin{figure*}
    \centering
    \includegraphics[width=0.95\textwidth]{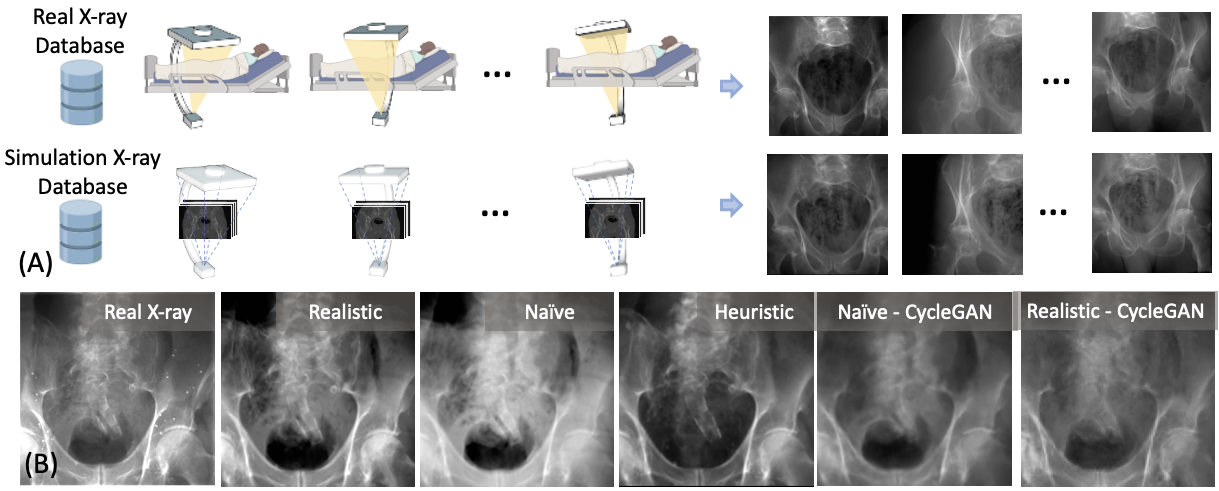}
    \caption{(A) Generation of precisely matched synthetic and real X-ray database: Real X-rays and CT scans are acquired from cadaveric specimens and registered to obtain the relative camera poses. Using these poses, synthetic X-rays can be generated from the CTs that precisely match the real X-ray data in all aspects but appearance. (B) Changes in (synthetic) X-ray appearance based on simulation paradigm.}
    \label{fig:matched_datase}
\end{figure*}

\subsection{Precisely Controlled Investigations on SyntheX' Sim2Real Performance Using the Hip Imaging Usecase}
Beyond presenting SyntheX for various clinical tasks, we present experiments on a unique dataset for hip imaging that enables the isolation of the effect that the domain gap has on simulation to reality ~(\emph{Sim2Real}) AI model transfer. On the task of anatomical landmark detection and anatomy segmentation in hip X-ray, we study the most commonly used domain generalization techniques, namely domain randomization and domain adaptation, and further consider different X-ray simulators, image resolution, and training dataset size. We introduce details on these experiments next.

\subsubsection{Precisely Matched Hip Dataset}
\label{sec:pelvic_dataset}
We created an accurately annotated dataset of 366 real hip fluoroscopic images and corresponding high-resolution CT scans of six lower torso cadaveric specimens with manual label annotations~\cite{grupp2020automatic}, which constitutes the basis of our unique dataset that enables precisely controlled benchmarking of domain shift. For each of the real X-ray images, the X-ray camera pose was accurately estimated using a comprehensive 2D/3D image registration pipeline~\cite{grupp2020automatic}. We then generated synthetic X-ray images (digitally reconstructed radiographs, DRRs) that precisely recreate the spatial configurations and anatomy of the real X-ray images and only differ in the realism of the simulation~(Fig.~\ref{fig:matched_datase}~(A)). Because synthetic images precisely match the real dataset, all labels in 2D and 3D apply equally. Details of the dataset creation are introduced in Section.~\ref{method:benchmark-dataset}.

We studied three different X-ray image simulation techniques, including na\"ive DRR generation, xreg DRR~\cite{grupp2019pose} and DeepDRR~\cite{unberath2018deepdrr,unberath2019enabling}, which we will refer to as \textit{Na\"ive}, \textit{Heuristic} and \textit{Realistic} simulations. They differ in the considerations of modeling realistic X-ray imaging physical effects. Fig.~\ref{fig:matched_datase}~(B) presents a comparison of image appearance between the different simulators and a corresponding real X-ray image. 

\subsubsection{Domain Randomization and Adaptation}
Domain randomization is a domain generalization technique that inflicts drastic changes on the input image appearances. This produces training samples with dramatically altered appearance, which forces the network to discover more robust associations between input image features and desired target. These more robust associations have been demonstrated to improve the generalization of machine learning models when transferred from one domain to another (here: from simulated to real X-ray images, respectively). We implemented two levels of domain randomization effects, namely regular domain randomization and strong domain randomization. Details are described in Section.~\ref{sec:dr_methods}.

Other than domain randomization which does not assume knowledge or sampling of the target domain at training time, domain adaptation techniques attempt to mitigate the domain gap's detrimental effect by aligning features across the source (training domain, here: simulated data) and target domain (deployment domain, here: real X-ray images). As such, domain adaptation techniques require samples from the target domain at training time. Recent domain adaptation techniques have increased the suitability of the approach for \emph{Sim2Real} transfer because they now allow for the use of unlabeled data in the target domain. We conducted experiments using two common domain adaptation methods: CycleGAN~\cite{zhu2017unpaired} and adversarial discriminative domain adaptation~(ADDA)~\cite{haq2020adversarial}. The two methods are similar in that they attempt to align properties of real and synthetic domains, and differ based on what properties they seek to align. While CycleGAN operates directly on the images, ADDA seeks to align higher-level feature representations, i.\,e., image features after multiple convolutional neural network layers. Example CycleGAN generated images are shown in Fig.~\ref{fig:matched_datase}~(B). More details of CycleGAN and ADDA training are provided in Section.~\ref{sec:da_methods}.



\subsection{Model and Evaluation Paradigm}
\label{sec:model_evaluation}

Because the focus of our experiments is to demonstrate convincing \emph{Sim2Real} performance, we rely on a well-established backbone network architecture, namely the U-Net~\cite{ronneberger2015u}, for all tasks. 
Segmentation networks for all clinical applications are trained to minimize the Dice loss~($L_{seg}$)~\cite{milletari2016v}, which evaluates the overlap between predicted and ground truth segmentation labels. 
For hip image analysis and surgical tool detection, we adjust the U-Net architecture as shown in Fig.~\ref{fig:unet_model} to concurrently estimate landmark locations. Reference landmark locations are represented as symmetric Gaussian distributions centered on the true landmark locations (zero when landmark is invisible). This additional prediction target is penalized using~($L_{ld}$), the mean squared error between network prediction and reference landmark heatmap.

\begin{Figure}
    \includegraphics[width=\textwidth]{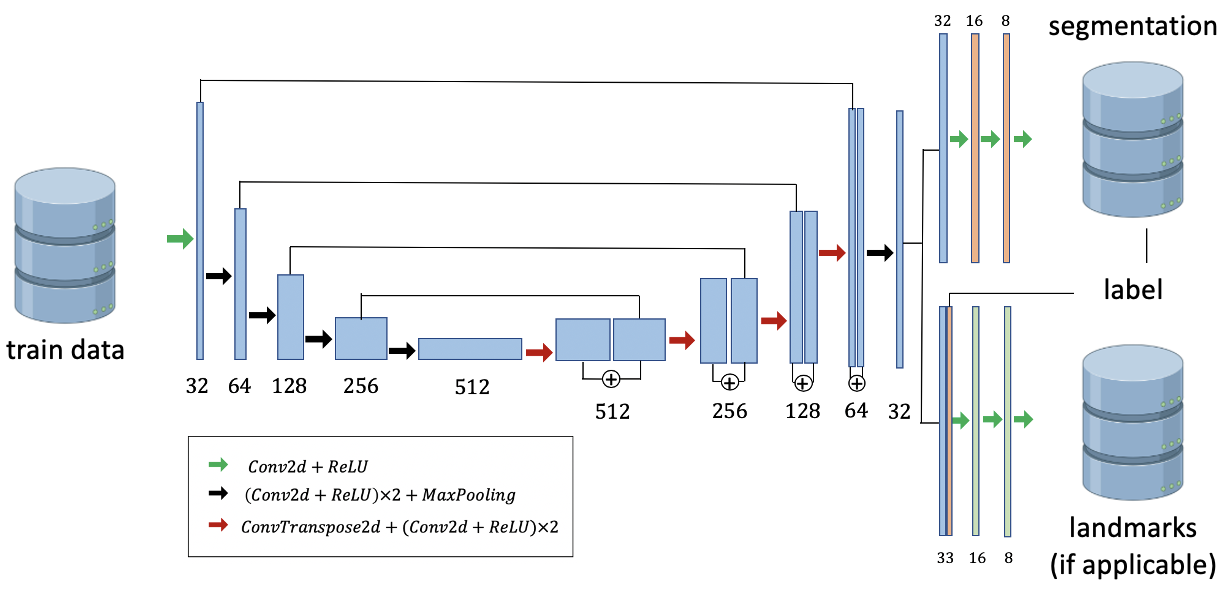}
    \captionof{figure}{U-net based concurrent segmentation and landmark detection network architecture for multi-task learning.}
    \label{fig:unet_model}
\end{Figure}

For evaluation purposes, we report the landmark accuracy as the $l_2$ distance between predicted and ground truth landmark positions. Further, we use the Dice score to quantitatively assess segmentation quality for hip imaging and surgical tool detection. The COVID-19 lesion segmentation performance is reported using confusion matrix metrics to enable comparison with prior work~\cite{wang2020noise}. 

For all three tasks, we report both \emph{Sim2Real} and \emph{Real2Real} performances. The \emph{Sim2Real} performance was computed on all testing real X-ray data. The \emph{Real2Real} experiments were conducted using k-fold cross-validation, and we report the performance as an average of all testing folds. For the hip imaging benchmark studies, we further carefully designed the evaluation paradigm in a leave-one-specimen-out fashion. For each experiment, the training and validation data consisted of all labeled images from all but one specimen while all labeled images from the remaining specimen were used as test data. The same data split was strictly preserved also for training of domain adaptation methods to avoid leakage and optimistic bias. On the scaled-up dataset, we used all synthetic images for training and evaluated on all real data in the benchmark dataset. 

A specially designed assessment curvature plot is used for reporting pelvic landmark detection performance. This way of measuring landmark detection performance provides detailed information on the two desirable attributes of such an algorithm: 1) completeness and 2) precision of detected landmarks. Landmarks are considered valid (activated) if their heatmap prediction is higher than a confidence threshold~($\phi$). The mean landmark detection error~($e^{ld}$) is reported as the average error over all activated landmarks. The ratio~($p$) of the activated landmarks over all landmarks is a function of $\phi$. Thus, we created plots to demonstrate the relationship between $e^{ld}$ and $p$, which shows the change of the error as we lower the threshold to activate more landmarks. Ideally, we would like a model to have a 0.0\,mm error with a 100\,\% activation percentage, corresponding to a measurement in the bottom right corner of the plots in Fig.~\ref{fig:landmarkPlots}. We selected the activation percentage of 90\% to report the numeric results for all ablation study methods in Table.~\ref{table-landmark}. 

\section{Results}
\subsection{Primary Findings}
We find that across all three clinical tasks, namely hip imaging, surgical robotic tool detection, and COVID-19 lesion segmentation, models trained using the SyntheX \emph{Sim2Real} model transfer paradigm when evaluated on real data perform comparably to or even better than models trained directly on real data. This finding suggests that Synthex, i.\,e., the realistic simulation of X-ray images from CT combined with domain randomization, is a feasible cost- and time-effective, and valuable approach to the development of learning-based X-ray image analysis algorithms that preserve performance during deployment on real data.

\subsubsection{Hip Imaging}
We present the multi-task detection results of hip imaging on images with $360\times 360$\,px in Table.~\ref{table-hip-imaging}. Both landmark detection and anatomical structure segmentation performance achieved using SyntheX \emph{Sim2Real} model transfer are superior than those of \emph{Real2Real} when considering averaged metrics. The \emph{Sim2Real} predictions are more stable with respect to their standard deviation: landmark error as 6.29\,mm, Dice score as 0.221, compared to \emph{Real2Real} with 15.30\,mm and 0.248, respectively. We attribute this improvement to the flexibility of the SyntheX approach, providing the possibility of simulating a richer spectrum of image appearances from more hip CT samples and varied X-ray geometries as compared to the limited data sourced from complex real-world experiments.

\begin{center}
\captionof{table}{Average performance metrics for hip imaging as a mean of 5-fold individual testing on 366 real hip X-ray images.}
\label{table-hip-imaging}
\begin{threeparttable}
    \begin{tabular}{c|cc}
\midrule\midrule
 & Landmark Error~(mm) & Dice Score\\
\hline
    \emph{Sim2Real} & $6.29\pm6.29$  & $0.818\pm0.221$ \\
    \emph{Real2Real}  & $8.15\pm15.30$ & $0.759\pm0.248$ \\
\bottomrule
\end{tabular}
\end{threeparttable}
\vspace{0.5em}
\end{center}

\subsubsection{Surgical Robotic Tool Detection}
Results of the surgical tool detection task are summarized in Table.~\ref{table-surgical-tool}. The landmark detection errors of \emph{Sim2Real} and \emph{Real2Real} are comparable with a mean localization accuracy of 2.13\,mm and 1.90\,mm, respectively. However, the standard deviation of \emph{Sim2Real} error is substantially smaller: 2.27\,mm versus 5.49\,mm. Further, with respect to segmentation Dice score, \emph{Sim2Real} outperforms \emph{Real2Real} by a large margin achieving a Dice score of 0.860 compared to 0.406, respectively. Overall, the results suggest that SyntheX is viable approach to developing deep neural networks for this task, especially when the robotic hardware is in prototypic stages.

\begin{center}
\captionof{table}{Average performance metrics for surgical tool detection as a mean of 5-fold individual testing on 264 real X-ray images of the continuum manipulator.}
\label{table-surgical-tool}
\begin{threeparttable}
    \begin{tabular}{c|cc}
\midrule\midrule
 & Landmark Error~(mm) & Dice Score\\
\hline
    \emph{Sim2Real} & $2.13\pm2.27$  & $0.860\pm 0.115$ \\
    \emph{Real2Real}  & $1.90\pm5.49$ & $0.406\pm 0.194$ \\
\bottomrule
\end{tabular}
\end{threeparttable}
\vspace{0.5em}
\end{center}

\subsubsection{COVID-19 Lesion Segmentation}
Results of COVID lesion segmentation are presented in Table.~\ref{table-covid}. The overall mean accuracy of SyntheX training reaches 85.03\% as compared to 93.95\% of real data training. The \emph{Sim2Real} performance is similar to \emph{Real2Real} in terms of sensitivity and specificity, but falls short in the other metrics. We attribute this deterioration to the inconsistency of the COVID lesion annotations between the training CT data and the real X-ray data. The results suggest that SyntheX is capable of handling soft tissue-based tasks, such as COVID-19 lesion segmentation. 

\begin{table*}
\caption {Average performance metrics (\%) for COVID-19 infected region segmentation as a mean of 5-fold individual testing on 2,951 real COVID-19 real chest X-ray images.}
\label{table-covid}
\centering
\begin{threeparttable}
    \begin{tabular}{c|cccccc}
\midrule\midrule
 & Sensitivity & Specifcity & Precision & F1-Score & F2-Score & Accuracy\\
\hline
    \emph{Sim2Real} & $76.49\pm17.25$  & $88.11\pm6.03$ & $48.46\pm27.49$ & $53.26\pm22.67$ & $61.43\pm25.53$ & $85.03\pm5.65$\\
    \emph{Real2Real}  & $81.37\pm17.75$ & $96.44\pm4.07$ & $73.67\pm25.84$ & $73.48\pm20.97$ & $76.90\pm25.95$ & $93.95\pm4.77$ \\
\bottomrule
\end{tabular}
\end{threeparttable}
\vspace{0.5em}
\end{table*}

\subsection{Sim2Real Benchmark Findings}
\label{sec:sim2real-benchmark}
Based on our precisely controlled hip imaging ablation studies including comparisons of 1) simulation environment, 2) domain randomization and domain adaptation effects, and 3) image resolution, we observed that training using realistic simulation with strong domain randomization performs on a par with models trained on real data or models trained on synthetic data but with domain adaptation, yet, \emph{does not require any real data at training time}. Training using realistic simulation consistently outperformed na\"ive or heuristic simulations. Training using scaled-up realistic simulation data with domain randomization achieved the best performance on this task, even outperforming real data-trained models due to the effectiveness of larger training data. Thus, realistic simulation of X-ray images from CT combined with domain randomization, which we refer to as the SyntheX model transfer concept, is a most promising approach to catalyze learning-based X-ray image analysis. The specially-designed landmark detection error metric plot, which summarizes the results across all ablations on images with $360\times360$\,px is shown in Fig.~\ref{fig:landmarkPlots}. We plotted the \emph{Real2Real} performance using golden curves as a baseline comparison to all the other ablation methods.

\subsubsection{The Effect of Domain Randomization}
\label{sec:dr_effect}
Across all experiments, we observed that networks trained with strong domain randomization consistently achieved better performance than those with regular domain randomization. This is expected because strong domain randomization introduces more drastic augmentations, which samples a much wider spectrum of possible image appearance and promotes the discovery of more robust features that are less prone to overfitting. The only exception is the training on na\"ively simulated images, where training with strong domain randomization results in much worse performance compared to regular domain randomization. We attribute this to the fact that the contrast of bony structures, which are most informative for the task considered here, are already much less pronounced in na\"ive simulations. Strong domain randomization then further increases problem complexity, to the point were performance deteriorates. 

\begin{figure*}
    \centering
    \includegraphics[width=\textwidth]{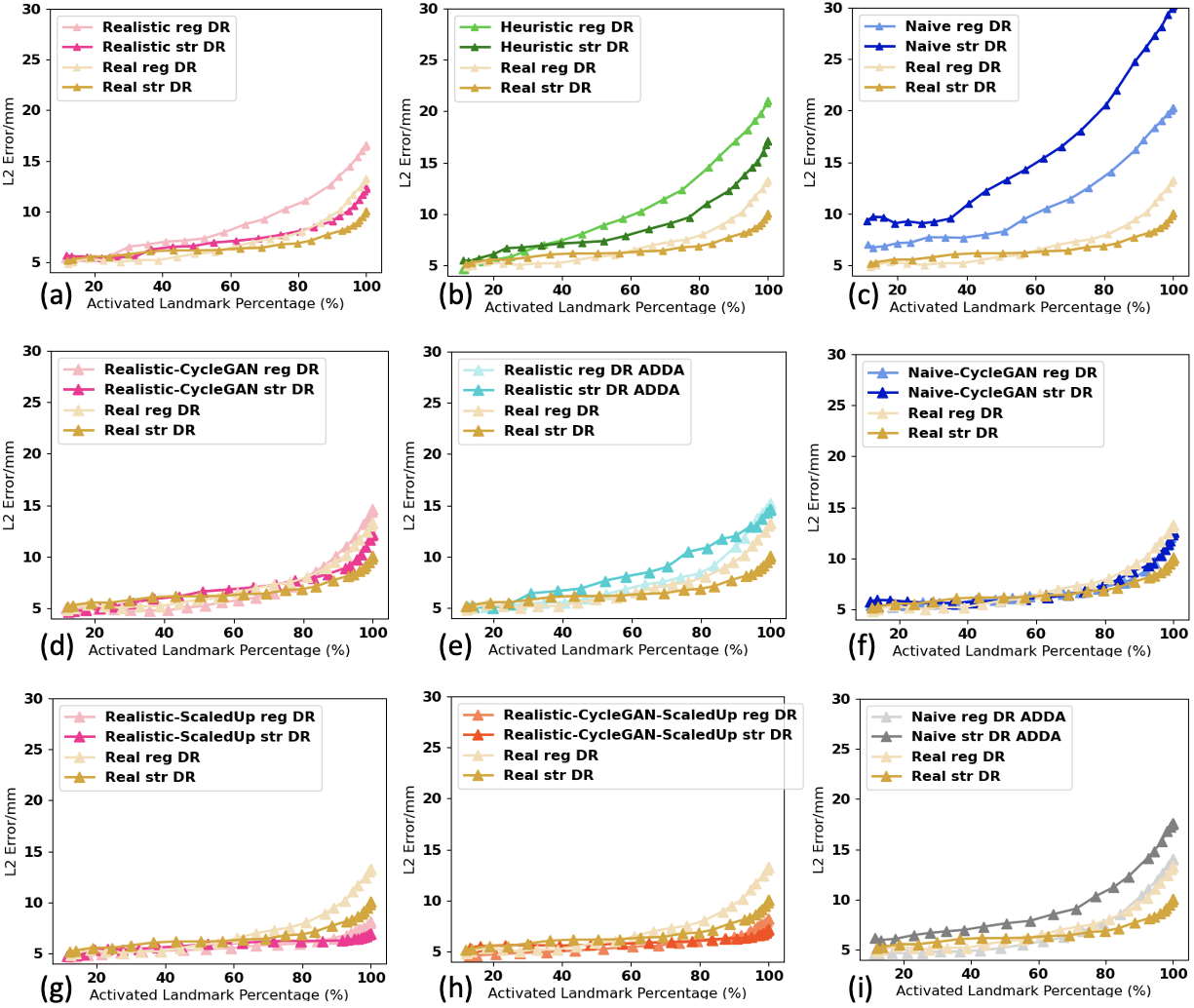}
    \caption{Plots of average landmark detection error curves with respect to activated landmark percentage. \emph{Real2Real} performance on the controlled dataset is shown in gold. An ideal curve should approach the bottom right corner: all landmarks detected with perfect localization. Each subplot compares the baseline \emph{Real2Real} performance curve to varied \emph{Sim2Real} methods that are evaluated on the same real data test set. The \emph{Sim2Real} technique of the specific method is identified in the upper left corner of each subplot. We use Real, Realistic, Heuristic and Na\"ive to refer to the image domains with decreasing level of realism, which are defined in Section~\ref{method:benchmark-dataset}. Domain names followed by ``CycleGAN'' mean the training data are generated using CycleGAN trained between the specific image domain and the real image domain. ``reg DR'' and ``str DR'' refer to regular domain randomization and strong domain randomization, respectively. ``ADDA'' refers to adversarial discriminative domain adaptation. (a)-(c) present performance comparison of methods trained on precisely matched datasets. (d)-(f)(i) further evaluates the added effect of using domain adaptation techniques again using precisely matched datasets. (g),(h) demonstrate improvements in \emph{Sim2Real} performance on the same real data test set when a larger, scaled-up synthetic training set is used. All the results correspond to input image size of $360\times 360$\,px.}
    \label{fig:landmarkPlots}
\end{figure*}

From Fig.~\ref{fig:landmarkPlots}(a)-(c), we see that realistic simulation~(DeepDRR) outperforms all other X-ray simulation paradigms in both regular domain randomization and strong domain randomization settings. Realistic simulation trained using strong domain randomization even outperforms \emph{Real2Real} with regular domain randomization. Since our experiments were precisely controlled and the only difference between the two scenarios is the image appearance due to varied simulation paradigm in the training set, this result supports the hypothesis that realistic simulation of X-rays using DeepDRR performs best for model transfer to real data. 

\subsubsection{The Effect of Domain Adaptation}
\label{sec:da_effect}
From Fig.~\ref{fig:landmarkPlots} (d)(f), we observe that both Realistic-CycleGAN and Na{\"i}ve-CycleGAN achieve comparable performance to \emph{Real2Real}. This means that images generated from synthetic images via CycleGAN have similar appearance, despite the synthetic training domains being different. The improvements over training purely on the respective synthetic domains~(Fig.~\ref{fig:landmarkPlots} (a)(c)) confirms that CycleGAN is useful for domain generalization. ADDA training also improves the performance over non-adapted transfer, but does not perform at the level of CycleGAN models. Interestingly, ADDA with strong DR shows deteriorated performance than regular DR~(Fig.~\ref{fig:landmarkPlots} (e)(i)). This is because the drastic and random appearance changes due to DR complicate domain discrimination, which in turn has adverse effects on overall model performance.   

\subsubsection{Scaling Up the Training Data}
We selected the best performing methods from the above domain randomization and domain adaptation ablations on the controlled dataset. These methods were realistic simulation with domain randomization and CycleGAN training based on realistic simulation, respectively, and trained on the scaled-up dataset, which contains much larger variety of anatomical shape and imaging geometry, i.\,e., synthetic C-arm poses. 

\begin{figure*}
    \centering
    \includegraphics[width=1.0\textwidth]{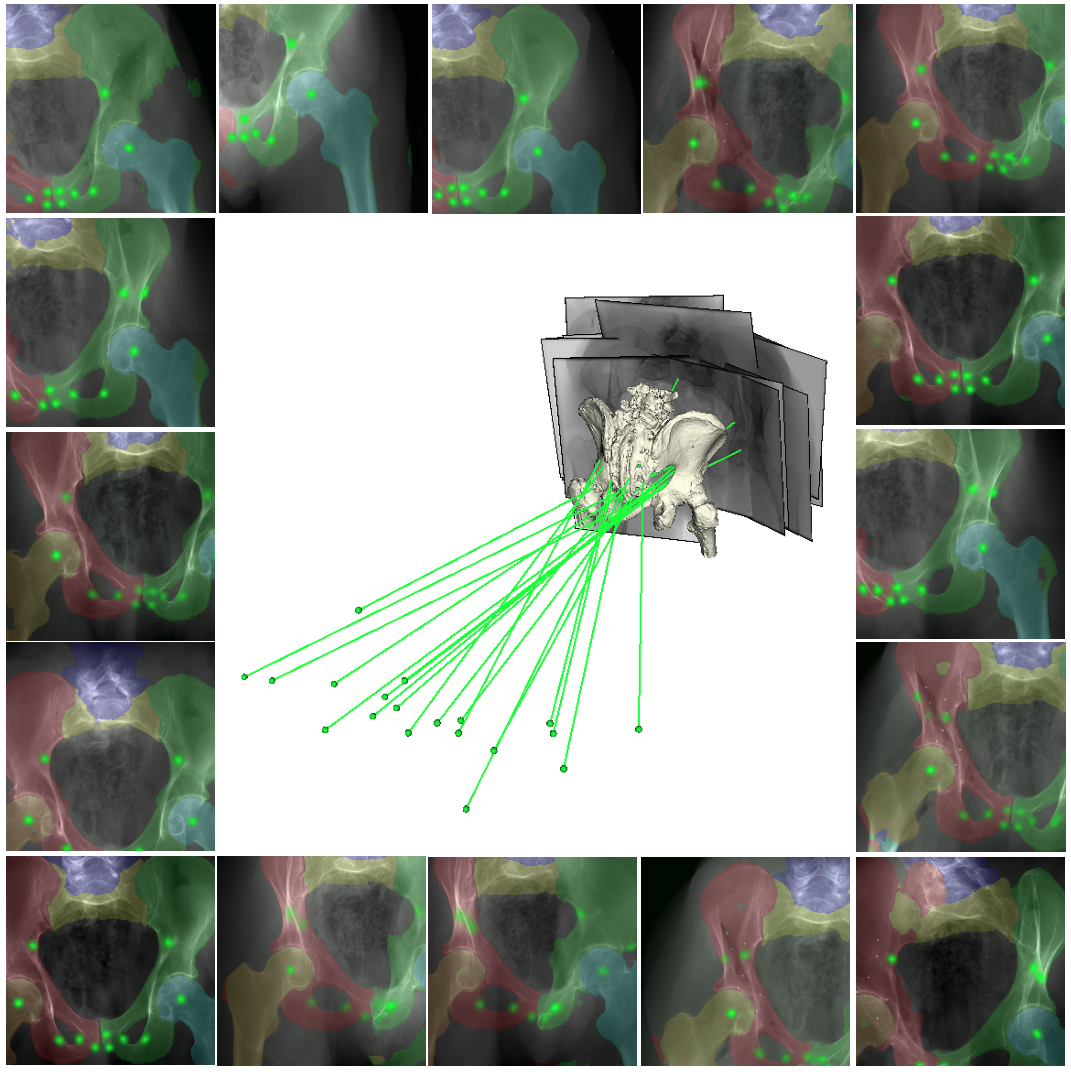}
    \caption{Qualitative results of the segmentation and landmark detection presented as overlays on testing data using the model trained with scaled up SyntheX data. Anatomical segmentation structures are blended with various RGB colors. Landmark heatmap responses are visualized in green. The projection geometries corresponding to the images relative to a 3D bone mesh model of the anatomy are presented in the center. The X-ray sources are shown as green dots and the principal rays are shown as green lines.}
    \label{fig:result_visualization}
\end{figure*}

With more training data and geometric variety, we found that all scaled-up experiments outperform the \emph{Real2Real} baseline on the benchmark dataset~(Fig.~\ref{fig:landmarkPlots}~(g)(h)). With 90\,\% of the landmarks activated, the model trained with strong domain randomization on realistically synthesized but large data (SyntheX, as reported above) achieved a mean landmark distance error of $6.29\pm6.29$\,mm. This is significantly better than the \emph{Real2Real} baseline (p=0.032), despite no real data being used for model training.  Fig.~\ref{fig:result_visualization} presents a collection of qualitative visualizations of the detection performance of this synthetic data-trained model when applied to real data. This result suggests that training with strong domain randomization and/or adaptation on large-scale, realistically synthesized data is a feasible alternative to training on real data. Training on large-scale data processed by CycleGAN achieved comparable performance~($6.59\pm7.25$~mm) as pure realistic simulation with domain randomization, but comes at the disadvantage that real data with sufficient variability must be available at training time to enable CycleGAN training.

\begin{table*}[t!]
\caption {Segmentation Dice Score as a mean of 5-fold individual testing on 366 real hip X-ray images. The Dice score ranges from 0 to 1, with larger values corresponding to better segmentation performance.}
\label{table-dice}
\centering
\begin{threeparttable}
    \begin{tabular}{c|cc|cc|cc}
\midrule\midrule
 Training Data & \multicolumn{2}{c|}{180x180} & \multicolumn{2}{c}{360x360} & \multicolumn{2}{|c}{480x480}\\
\cline{2-7}
Domain & regular DR & strong DR & regular DR & strong DR & regular DR & strong DR\\
\hline
    RealXray~(\emph{Real2Real}) & $0.775\pm0.235$  & $0.784\pm0.214$ & $0.783\pm0.232$ & $0.759\pm0.248$ & $0.739\pm0.266$ & $0.751\pm0.265$\\
    Realistic  & $0.730\pm0.240$ & $0.787\pm0.211$ & $0.751\pm0.241$ & $0.760\pm0.250$ & $0.720\pm0.256$ & $0.700\pm0.279$ \\
    Heuristic & $0.669\pm0.273$ & $0.737\pm0.249$ & $0.683\pm0.265$ & $0.682\pm0.286$ & $0.655\pm0.277$ & $0.668\pm0.298$\\
    Na\"ive & $0.689\pm0.256$ & $0.680\pm0.278$ & $0.687\pm0.266$ & $0.572\pm0.309$ & $0.653\pm0.278$ & $0.578\pm0.305$\\
    Realistic-Cyc & $0.778\pm0.217$ & $0.778\pm0.220$ & $0.760\pm0.248$ & $0.733\pm0.267$  & $0.741\pm0.255$ & $0.688\pm0.291$\\
    Na\"ive-Cyc & $0.784\pm0.198$ & $0.750\pm0.230$ & $0.739\pm0.254$ & $0.736\pm0.258$ & $0.726\pm0.254$ & $0.673\pm0.292$\\
    Realistic-ADDA & $0.767\pm0.224$ & $0.754\pm0.231$ & $0.726\pm0.292$ & $0.731\pm0.256$ & $0.704\pm0.279$ & $0.727\pm0.256$\\
    Na\"ive-ADDA & $0.491\pm0.405$ & $0.678\pm0.266$ & $0.693\pm0.297$ & $0.662\pm0.265$ & $0.693\pm0.273$ & $0.592\pm0.306$\\
    Realistic-Scaled & $\mathbf{0.857\pm0.184}$ & $\mathbf{0.853\pm0.179}$ & $\mathbf{0.838\pm0.221}$ & $\mathbf{0.818\pm0.221}$ & $0.783\pm0.262$ & $\mathbf{0.823\pm0.221}$\\
    Realistic-Cyc-Scaled & $0.822\pm0.213$ & $0.794\pm0.232$ & $0.824\pm0.225$ & $0.789\pm0.240$ & $\mathbf{0.789\pm0.241}$ & $0.812\pm0.227$\\
\bottomrule
\end{tabular}
\end{threeparttable}
   \begin{tablenotes}
     \item Note: The best performance results are bolded. Training/testing image resolutions are listed in the top row. DR is short for domain randomization. Cyc is short for CycleGAN. ADDA refers to adversarial discriminative domain adaptation. ``-Scaled'' means training on the scaled-up dataset. 
   \end{tablenotes}
\vspace{0.5em}
\end{table*}

\begin{table*}[t!]
  \caption{Landmark Detection Errors~(mm) at 90\% activation percentage as a mean of 5-fold individual testing on 366 real hip X-ray images. Lower values are better.}
  \label{table-landmark}
  \centering 
  \begin{threeparttable}
    \begin{tabular}{c|cc|cc|cc}
     \midrule\midrule
    Training Data & \multicolumn{2}{c|}{180x180} & \multicolumn{2}{c}{360x360} & \multicolumn{2}{|c}{480x480}\\
     \cline{2-7}
    Domain & regular DR & strong DR & regular DR & strong DR & regular DR & strong DR\\
\hline
    RealXray~(\emph{Real2Real}) & $9.98\pm22.58$  & $7.78\pm11.94$ & $8.93\pm19.76$ & $8.15\pm15.30$ & $8.98\pm21.38$ & $7.59\pm16.12$\\
    Realistic & $14.33\pm32.61$ & $8.41\pm14.47$ & $12.62\pm27.68$ & $9.05\pm19.37$ & $14.96\pm34.45$ & $13.06\pm26.52$ \\
    Heuristic & $20.84\pm44.16$ & $10.69\pm22.92$ & $14.54\pm32.88$ & $12.25\pm26.82$ & $17.59\pm39.55$ & $12.85\pm29.49$\\
    Na\"ive & $13.12\pm30.87$ & $13.03\pm21.20$ & $16.22\pm39.71$ & $20.55\pm35.92$ & $18.53\pm40.88$ & $19.46\pm37.48$\\
    Realistic-Cyc & $8.65\pm19.84$ & $8.19\pm13.55$ & $8.57\pm18.69$ & $8.90\pm17.47$  & $10.78\pm26.9$ & $8.75\pm17.4$\\
    Na\"ive-Cyc & $8.56\pm18.46$ & $9.05\pm10.75$ & $7.63\pm16.10$ & $9.29\pm18.73$ & $9.14\pm21.82$ & $11.73\pm25.06$\\
    Realistic-ADDA & $11.19\pm24.80$ & $11.43\pm25.83$ & $11.01\pm24.69$ & $12.92\pm23.76$ & $16.42\pm37.61$ & $9.24\pm17.53$\\
    Na\"ive-ADDA & $7.90\pm17.56$ & $11.84\pm25.63$ & $10.42\pm25.05$ & $14.17\pm30.89$ & $16.62\pm40.12$ & $22.88\pm41.53$\\
    Realistic-Scaled & $\mathbf{5.91\pm8.43}$ & $\mathbf{6.06\pm7.10}$ & $\mathbf{6.80\pm9.53}$ & $\mathbf{6.29\pm6.29}$ & $8.12\pm19.35$ & $5.99\pm12.19$\\
    Realistic-Cyc-Scaled & $6.79\pm9.70$ & $6.87\pm9.58$ & $6.87\pm13.19$ & $6.59\pm7.25$ & $\mathbf{6.43\pm13.67}$ & $\mathbf{5.52\pm4.85}$\\
     \bottomrule
    \end{tabular}
  \end{threeparttable}
  \begin{tablenotes}
     \item Note: The best performance result is bolded. Training/testing image resolutions are listed in the top row. DR is short for domain randomization. Cyc is short for CycleGAN. ADDA refers to adversarial discriminative domain adaptation. ``-Scaled'' means training on scaled-up dataset. 
   \end{tablenotes}
\end{table*}

\section{Discussion}
We present the general use cases of SyntheX on various scenarios, including purely bony anatomy~(the hip), metallic artificial surgical tool and soft tissue~(lung COVID lesion). Our experiments on three varied clinical tasks demonstrate that the performance of models trained using SyntheX - on real data - meets or exceeds the performance of real data-trained models. We show that generating realistic synthetic data is a viable resource for developing machine learning models compared to collecting largely annotated real clinical data.

The hip imaging ablation experiments reliably quantify the effect of the domain gap on real data performance for varied \emph{Sim2Real} model transfer approaches. This is because all aleatoric factors that usually confound such experiments are precisely controlled for, with alterations to image appearances due to the varied image simulation paradigms being the only source of mismatch. The aleatoric factors we controlled for include anatomy, imaging geometries, ground truth labels, network architectures, and hyperparameters. In particular, the viewpoints and 3D scene recreated in simulation were identical to the real images, which to our knowledge has not yet been achieved. From these results, we draw the following conclusions: 
\begin{itemize}
    \item Physics-based, realistic simulation of training data using the DeepDRR framework results in models that generalize better to the real data domain compared to models trained on less realistic, i.\,e., na\"ive or heuristic simulation paradigms. This suggests, not surprisingly, that matching the real image domain as closely as possible directly benefits generalization performance.
    \item Realistic simulation combined with strong domain randomization (SyntheX) performs on a par with both, the best domain adaptation method (CycleGAN with domain randomization) as well as real data training. However, because SyntheX does not require any real data at training time, this paradigm has clear advantages over domain adaptation. Specifically, it saves the effort to acquire real data early in development or to design additional machine learning architectures that perform adaptation. This makes SyntheX particularly appealing for the development of novel instruments or robotic components, real images of which can simply not be acquired early during conceptualization.
\end{itemize}
Realistic simulation using DeepDRR is as computationally efficient as na\"ive simulation, both of which are orders of magnitude faster than Monte Carlo simulation~\cite{unberath2018deepdrr}. Further, realistic simulation using DeepDRR brings substantial benefits in regards to \emph{Sim2Real} performance and self-contained data generation and training. These findings are very encouraging and strongly support the hypothesis that training on synthetic radiographs simulated from 3D CT is a viable alternative to real data training, or at a minimum, a strong candidate for pre-training.

Compared to acquiring real patient data, generating large-scale simulation data is more flexible, timely efficient, low-cost and avoids privacy concerns. For the hip image analysis use case, we performed experiments based on 10,000 synthetic images from 20 hip CT scans. Training with realistic simulation and strong domain randomization outperformed \emph{Real2Real} training at the 90\,\% activation level but generally improved performance as seen by a flattened activation vs. error curve (Fig.\ref{fig:landmarkPlots}~(g)). The performance of training with CycleGAN with larger datasets was similar. These findings suggest that scaling-up data for training is an effective strategy to improve performance both in- and outside of the training domain. Scaling up training data is costly or impossible in real settings, but in comparison is easily possible using data synthesis. Having access to more varied data samples during training helps the network parameter optimization find a more stable solution, that also transfers better.

We have found that \emph{Sim2Real} model transfer performs best for scenarios where real data and corresponding annotations are particularly hard to obtain. This is evidenced by the change in performance gap between \emph{Real2Real} and \emph{Sim2Real} training, where \emph{Sim2Real} performs particularly well for scenarios where little real data is available, such as for hip imaging and robot tool detection, and hardly matches \emph{Real2Real} performance for use cases where abundant real data exists, such as COVID-19 lesion segmentation. The value of SyntheX thus primarily derives from the possibility of generating large synthetic training datasets for innovative applications, e.\,g., including custom-designed hardware~\cite{gao2019localizing,gao2021fluoroscopic} or novel robotic imaging paradigms~\cite{zaech2019learning,thies2020learning}, the data for which could not otherwise be obtained. Second, SyntheX can complement real datasets by providing synthetic samples that exhibit increased variability in anatomies, imaging geometries, or scene composition. Finally, the SyntheX simulation paradigm enables generation of precise annotations, e.\,g., the lesion volume in the COVID-19 use case, that could not be derived otherwise. 

Interestingly, although domain adaptation techniques (CycleGAN and ADDA) have access to data in the real domain, these methods outperformed domain generalization techniques (here, domain randomization) only by a small margin in the controlled study. The performance of ADDA training heavily depends on the choices of additional hyperparameters, such as the design of the discriminator, number of training cycles between task and discriminator network updates, and learning rates, among others. Thus, it is non-trivial to find the best training settings, and these settings are unlikely to apply to other tasks. Because CycleGAN performs image-to-image translation, a complicated task, it requires sufficient and sufficiently diverse data in the real domain to avoid overfitting. Further, using CycleGAN requires an additional training step of a large model, which is memory intensive and generally requires long training time. In certain cases, CycleGAN models could also introduce undesired effects. Jaihyun et al. find that performance of CycleGAN is highly dependent on the dataset, potentially resulting in unrealistic images with less information content than the original images~\cite{park2019adaptive}. Finally, because real domain data is being used in both domain adaptation paradigms, adjustments to the real data target domain, e.\,g., use of a different C-arm X-ray imaging device or design changes to surgical hardware, may require de novo acquisition of real data and re-training of the models. In contrast, SyntheX resembles a plug-and-play module, to be integrated into any learning-based medical imaging tasks, which is easy to set up and use. Similar to multi-scale modeling~\cite{alber2019integrating} and in silico virtual clinical trials~\cite{viceconti2016silico,badano2018evaluation}, SyntheX has the potential to envision, implement, and virtually deploy solutions for image-guided procedures and evaluate their potential utility and adequacy. This makes SyntheX a promising tool that may replicate traditional development workflows solely using computational tools.

Despite the promising outlook, our study has several limitations. 
First, while the real X-ray and CT data of cadaveric specimens used for the hip imaging and robotic tool segmentation task is of respectable size for this type of application, it is small compared to some dataset sizes in general computer vision applications. However, the effort, facilities, time, and therefore, costs required to acquire and annotate a dataset of even this size are substantial due to the nature of the data. To counter this shortcoming, we performed leave-one-subject-out cross validations for all experiments and found that the clinical task of anatomical segmentation and landmark detection could be solved at an acceptable level that is comparable to previous studies on larger data~\cite{bier2018x,bier2019learning}. 

Second, the performance we report is limited by the quality of the CT and annotations. The spatial resolution of CT scans~(between 0.5 to 1.0\,mm in hip imaging and surgical robot tool segmentation, between 1.0 to 2.0\, mm in COVID lesion segmentation, isotropic) imposed a limitation on the resolution that can be achieved in 2D simulation. Pixel sizes of conventional detectors are as small as 0.2\,mm, smaller than the highest resolution scenario considered here. However, contemporary computer vision algorithms for image analysis tasks have only considered downsampled images in the ranges described here. Another issue arises from annotation mismatch, especially when annotations are generated using different processes for SyntheX training and evaluation on real 2D X-ray images. This challenge arose specifically in the COVID-19 lesion segmentation task, where 3D lesion labels generated from the pre-trained lesion segmentation network and used for SyntheX training are not consistent with the annotations on real 2D X-ray data. This is primarily for two reasons: First, because CT scans and chest X-ray images were not from the same patients, COVID-19 disease stages and extent of lesions are varied; second, because real chest X-rays were annotated in 2D smaller or more opaque parts of COVID-19 lesions may have been missed due to the projective and integrative nature of X-ray imaging. This mismatch in ground truth definition is unobserved but establishes an upper bound on the possible \emph{Sim2Real} performance. Further, realism of simulation can be improved with higher quality CT scans, super-resolution techniques, and advanced modeling techniques to more realistically represent anatomy at higher resolutions.

\section{Conclusion}
In this paper, we demonstrated that realistic simulation of image formation from human models combined with domain generalization or adaptation techniques is a viable alternative to large scale real data collection. We demonstrate its utility on three variant clinical tasks, namely hip imaging, surgical robotic tool detection and COVID lesion segmentation. Based on controlled experiments on a pelvic X-ray dataset, which is precisely reproduced in varied synthetic domains, we quantified the effect of simulation realism and domain adaptation/generalization techniques on \emph{Sim2Real} transfer performance. We found promising \emph{Sim2Real} performance of all models that were trained on realistically simulated data. The specific combination of training on realistic synthesis and strong domain randomization, which we refer to as SyntheX, is particularly promising. SyntheX-trained models perform on a par with real-trained models, making realistic simulation of X-ray-based clinical workflows and procedures a viable alternative or complement to real data acquisition. Because SyntheX does not require real data at training time, it is particularly promising for the development of machine learning models for novel clinical workflows or devices, including surgical robotics, before these solutions exist physically.

\section{Methods}
We introduce further details on the domain randomization and domain adaptation methods applied in our studies. We then provide additional information on experimental setup and network training details of the clinical tasks and benchmark investigations.

\subsection{Domain Randomization}
\label{sec:dr_methods}
Domain randomization effects were applied to the input images during network training. We studied two domain randomization levels: regular and strong domain randomization. Regular domain randomization included the most commonly used data augmentation schemes. For strong domain randomization, we included more drastic effects and combined them together. We use $x$ to denote a training image sample. The domain randomization techniques we introduced are as follows:

Regular domain randomization included: 1)~\textit{Gaussian noise injection}: $x+N(0,\sigma)$, where $\sigma$ was uniformly chosen from the interval $(0.005, 0.1)$ multiplied by the image intensity range. 2)~\textit{Gamma transform}: $norm(x)^\gamma$, where $x$ was normalized by its maximum and minimum value and $\gamma$ was uniformly selected from the interval $(0.7, 1.3)$. 3)~\textit{Random crop} $x$ was cropped at random locations using a square shape which has the dimension of 90\% $x$ size. Regular domain randomization methods were applied to every training iteration.

Strong domain randomization included: 1)~\textit{Inverting}: $max(x)-x$, where the maximum intensity value was subtracted from all image pixels. 2)~\textit{Impulse/Pepper/Salt noise injection}: 10\% of pixels in $x$ were replaced with one type of noise including impluse, pepper and salt. 3)~\textit{Affine transform}: a random 2D affine warp including translation, rotation, shear and scale factors was applied. 4)~\textit{Contrast}: $x$ was processed with one type of the contrast manipulations including linear contrast, log contrast and sigmoid contrast. 5)~\textit{Blurring}: $x$ was was processed with a blurring method including Gaussian blur~$N(\mu=0, \sigma=3.0)$ and average blur~(kernel size between $2\times2$ and $7\times7$). 6)~\textit{Box corruption}: a random number of box regions were corrupted with large noise. 7)~\textit{Dropout}: Either randomly dropped 1-10\% of pixels in $x$ to zero, or dropped them in a rectangular region with 2-5\% of the image size. 8)~\textit{Sharpening and embossing}: Sharpen $x$ blended the original image with a sharpened version with an alpha between 0 and 1~(no and full sharpening effect). Embossing added the sharpened version rather than blending it. 9) One of the pooling methods was applied to $x$: average pooling, max pooling, min pooling and median pooling. All of the pooling kernel sizes were between $2\times2$ and $4\times4$. 10)~\textit{Multiply}: Either changed brightness or multiplied $x$ element wise with 50-150\% of original value. 11)~\textit{Distort}: Distorted local areas of $x$ with a random piece-wise affine transformation. For each image, we still applied basic domain randomization but only randomly concatenated up to two strong domain randomization methods during each training iteration to avoid too heavy augmentation. 

\subsection{Domain Adaptation}
\label{sec:da_methods}
We select the two most commonly used domain adaptation approaches for our comparison study, which are CycleGAN~\cite{zhu2017unpaired} and adversarial discriminative domain adaptation~(ADDA)~\cite{haq2020adversarial}. CycleGAN was trained using unpaired synthetic and real images prior to task network training. All synthetic images were then processed with trained CycleGAN generators, to alter their appearance to match real data. We strictly enforced the data split used during task-model training so that images from the test set were excluded during both CycleGAN and task network training. ADDA introduced an adversarial discriminator branch as an additional loss to discriminate between features derived from synthetic and real images. We followed the design of Haq et al. to build the discriminator for ADDA training on the task of semantic segmentation~\cite{haq2020adversarial}. Both CycleGAN and ADDA models were tested using realistic and na{\"i}ve simulation images.

\subsubsection{CycleGAN} 
CycleGAN was applied to learn mapping functions between two image domains $X$ and $Y$ given training samples $\{x_i\}_{i=1}^N$ where $x_i \in X$ and $\{y_j\}_{j=1}^M$ where $y_j \in Y$. The model includes two mapping functions $G: X\rightarrow Y$ and $F: Y\rightarrow X$, and two adversarial discriminators $D_X$ and $D_Y$. The objective contains two terms: \emph{adversarial loss} to match the distribution between generated and target image domain; and \emph{cycle consistency loss} to ensure learned mapping functions are cycle-consistent.\\ 
For one mapping function $G: X\rightarrow Y$ with its discriminator $D_Y$,  the first term, \emph{adversarial loss}, can be expressed as:
\begin{align}
    \mathcal{L}_{\text{GAN}}(G,D_Y,X,Y) =& \ \mathbb{E}_{y \sim p_{\text{data}}(y)}[\log D_Y(y)] \nonumber \\
   +& \ \mathbb{E}_{x \sim p_{\text{data}}(x)}[\log (1-D_Y(G(x))],
\end{align}
where $G$ generates images $G(x)$ with an appearance similar to images from domain $Y$, while $D_Y$ tries to distinguish between translated samples $G(x)$ and real samples $y$. Overall, $G$ aims to minimize this objective against an adversary $D$ that tries to maximize it.
Similarly, there is an \emph{adversarial loss} for the mapping function $F:Y\rightarrow X$ with its discriminator $D_X$.\\

The second term, \emph{cycle consistency loss}, can be expressed as:
\begin{align}
    \mathcal{L}_{\text{cyc}}(G, F) =  & \ \mathbb{E}_{x\sim p_{\text{data}}(x)}\left[\norm{F(G(x))-x}_1\right] \nonumber \\
    + &\ \mathbb{E}_{y\sim p_{\text{data}}(y)}\left[\norm{G(F(y))-y}_1\right],
\end{align}
where for each image $x$ from domain $X$, $x$ should be recovered after one translation cycle, i.\,e., $x \rightarrow G(x) \rightarrow F(G(x)) \approx x$. Similarly, each image $y$ from domain $Y$ should be recovered as well. Zhu et al.~\cite{zhu2017unpaired} argued that learned mapping functions should be cycle-consistent to further reduce the space of possible mapping functions. The above formulation using domain discrimination and cycle consistency enables unpaired image translation, i.\,e., learning the mappings $G(x)$ and $F(y)$ without corresponding images.\\

The overall objective for CycleGAN training is expressed as:
\begin{align}
     \mathcal{L}(G,F,D_X,D_Y) = & \mathcal{L}_{\text{GAN}}(G,D_Y,X,Y) \nonumber \\
    +&\ \mathcal{L}_{\text{GAN}}(F,D_X,Y,X) \nonumber \\
    +& \  \lambda \mathcal{L}_{\text{cyc}}(G, F),
\end{align}

where $\lambda$ controls the relative importance of cycle consistency loss, aiming to solve:
\begin{equation}
    G^*,F^* = \arg\min_{G,F}\max_{D_x,D_Y} \mathcal{L}(G, F, D_X, D_Y).
\end{equation}

For the generator network, $6$ blocks for $128\times128$ images and $9$ blocks for $256\times 256$ and higher-resolution training images were used with instance normalization. For the discriminator network, a $70\times 70$ PatchGAN was used.

\subsubsection{Adversarial Discriminative Domain Adaptation} 
We applied the idea of Haq et al. on our pelvis segmentation and landmark localization task~\cite{haq2020adversarial}. The architecture consists of three components, including \textit{Segmentation and Localization network}, \textit{Decoder} and \textit{Discriminator}. The input to \textit{Segmentation and Localization network} is image~($x$) and the output prediction feature is $z$. The loss is $L_{seg}$ and $L_{ld}$ as introduced in Section~\ref{sec:clinical-task}. The \textit{Decoder} shared the same U-Net architecture, takes $z$ as input and the output is the reconstruction $R(z)$. The reconstruction loss, $L_{recons}$, is the mean squared error between $x$ and $z$. The \textit{Discriminator} was trained using an adversarial loss:
\begin{equation}
    \begin{aligned}
    L_{dis}(z) = & -\frac{1}{H\times W}\sum_{h, w}s\cdot log(D(z))\\
                 & + (1-s)\cdot log(1-D(z)),
    \end{aligned}
\end{equation}
where $H$ and $W$ are the feature dimension of the discriminator output, $s=0$ when $D$ takes target domain prediction~($Y_t$) as input, and $s=1$ when taking source domain prediction~($Y_s$) as input. The $Discriminator$ contributed an adversarial loss during training in order to bring in domain transfer knowledge. The adversarial loss is defined as:
\begin{equation}
    L_{adv}(x_t)=-\frac{1}{H\times W}\sum_{h, w}log(D(z_t)).
\end{equation}
Thus, the total training loss can be written as:
\begin{equation}
    \begin{aligned}
    L_t(x_s, x_t) = & L_{seg}(x_s) + L_{ld}(x_s) + \lambda_{adv}L_{adv}(x_t)\\
                    & + \lambda_{recons}L_{recons}(x_t),
    \end{aligned}
\end{equation}
where $\lambda_{adv}$ and $\lambda_{recons}$ are weight hyperparameters, which are empirically chosen to be 0.001 and 0.01, as suggested by Haq et al.~\cite{haq2020adversarial}.

\subsection{Clinical Tasks Experimental Details}
The SyntheX simulation environment was set up to approximate a Siemens CIOS Fusion C-Arm, which has image dimensions of $1536\times1536$, isotropic pixel spacing of 0.194\,mm/pixel, a source-to-detector distance of 1020\,mm, and a principal point at the center of the image.

\subsubsection{Hip Imaging}
Synthetic hip X-rays were created using 20 CT scans from the New Mexico Decedent Image Database~\cite{newmexico}. During simulation, we uniformly sampled the CT volume rotation in $[-45^\circ, 45^\circ]$, and translation Left/Right in [-50\,mm, 50\,mm], Interior/Superior in [-20\,mm, 20\,mm], Anterior/Posterior in [-100\,mm, 100\,mm]. We generated 18,000 images for training and 2,000 images for validation. Ground truth segmentation and landmark labels were projected from 3D using the projection geometry. 

We consistently trained the model for 30 epochs and selected the best validation model for evaluation. Strong domain randomization was applied at training time~(Section.~\ref{sec:dr_methods}). During evaluation, a threshold of 0.5 was used for segmentation and the landmark prediction was selected using the highest heatmap response location.

\subsubsection{Robotic Surgical Tool Detection}
We created 100 voxelized models of the continuum manipulator~(CM) in various configurations by sampling its curvature control point angles form a Gaussian distribution $N(\mu=0, \sigma=2.5^\circ)$. The CM base pose was uniformly sampled LAO/RAO in $[-30^\circ, 30^\circ]$, CRAN/CUAD in $[-10^\circ, 10^\circ]$, source-to-isocenter distance in [600\,mm, 900\,mm], and translation in x, y axes following a Gaussian distribution $N(\mu=0\,\text{mm}, \sigma=10\,\text{mm})$. We created DeepDRR synthetic images by projecting randomly selected hip CT scan from the 20 New Mexico Decedent Image Database CT scans used for hip imaging together with the CM model, which include 28,000 for training and 2,800 for validation. Ground truth segmentation and landmark labels were projected following each simulation geometry.

The network training details are in Section.~\ref{sec:network-details}, and strong domain randomization was applied~(Section.~\ref{sec:dr_methods}). The network was trained for 10 epochs and the best validation model was selected for evaluation. The performance was evaluated on 264 real CM X-ray images with manual ground truth label annotations. During evaluation, a threshold of 0.5 was used for segmentation and the landmark prediction was selected using the highest heatmap response location. The network training and evaluation routines are the same for the \emph{Real2Real} 5-fold training and testing.

\subsubsection{COVID-19 Lesion Segmentation}
We used 81 high quality CT scans from ImagEng lab~\cite{zaffino2021open} and 62 CT scans with resolution less than 2\,mm from UESTC~\cite{wang2020noise}, all diagnosed as COVID-19 cases, to generate synthetic chest X-ray data. The 3D lesion segmentations of CTs from ImagEng lab were generated using the pre-trained COPLE-Net~\cite{wang2020noise}. During DeepDRR simulation, we uniformly sampled the view pose of CT scans, rotation from $[-5^\circ, 5^\circ]$ in all three axes and source-to-isocenter distance in [350\,mm, 650\,mm], resulting in 18,000 training images and 1,800 validation images with a resolution of $224\times 224$\,px. A random shearing transformation from $[-30^\circ, 30^\circ]$ was applied on the CT scan and segmentations were obtained with a threshold of 0.5 on the predicted response. The corresponding lesion mask was projected from the 3D segmentation using the simulation projection geometry. 

The network training setups follow the descriptions in Section.~\ref{sec:network-details}. Strong domain randomization was applied during training time~(Section.~\ref{sec:dr_methods}). We trained the network for 10 epochs and selected the best model with the highest validation score for testing. The performance was evaluated on 2,951 real COVID-19 benchmark dataset~\cite{degerli2021covid}. During evaluation, the network segmentation mask was created using a threshold value of 0.5 on the original prediction. The network training and evaluation routines are the same for the \emph{Real2Real} 5-fold training and testing.

\subsubsection{Benchmark Hip Imaging Investigation}
\label{method:benchmark-dataset}
For every X-ray image, ground truth X-ray camera pose relative to the CT scan were estimated using an automatic intensity-based 2D/3D registration of the pelvis and both femurs~\cite{grupp2020automatic}. Every CT was annotated with segmentation of anatomical structures and anatomical landmark locations defined in Fig.~\ref{fig:clinical_task} (A). 2D labels for every X-ray image were then generated automatically by forward projecting the reference 3D annotations using the corresponding ground truth C-arm pose.

We generated synthetic data using three DRR simulators: Na\"ive DRR, xreg DRR and DeepDRR. Na\"ive DRR generation amounts to simple ray-casting and does not consider any imaging physics. This amounts to the assumption of a mono-energetic source, single material objects, and no image corruption, e.\,g., due to noise or scattering. Heuristic simulation~(xreg DRR\footnote{https://github.com/rg2/xreg}) performs a linear thresholding of the CT Hounsfield Units~(HU) to differentiate materials between air and anatomy prior to ray-casting. While this results in a more realistic appearance of the resulting DRRs, in that the tissue contrast is increased, the effect does not model imaging physical effects. Realistic simulation~(DeepDRR) simulates imaging physics by considering the full spectrum of the X-ray source, and relies on machine learning for material decomposition and scatter estimation. It also considers both signal dependent noise as well as readout noise together with detector saturation.

\subsection{Network Training Details}
\label{sec:network-details}
We used Pytorch for all implementations and trained the networks from scratch, using stochastic gradient descent with a learning rate of 0.0002, Nesterov momentum of 0.9, weight decay of 0.00001, and a constant batch size of five images. The multi-task network training loss is equally weighted between landmark detection loss and segmentation loss. For all ablation experiments in the precisely controlled study, we kept the same learning rate for the first 100 epochs and linearly decayed the rate to zero over the next 100 epochs.

\section*{Code Availability}
An updated repository for DeepDRR is available at \url{https://github.com/arcadelab/deepdrr}. The xReg registration software module is at \url{https://github.com/rg2/xreg}. The multi-task deep network module will be made publicly available on github upon publication.

\section*{Data Availability}
We provide access web links for public data used in our study:

The hip imaging CT scans are selected from the New Mexico Decedent Image Database: \url{https://nmdid.unm.edu/}.

The hip imaging real cadaveric CT scans and X-rays can be accessed here: \url{https://github.com/rg2/DeepFluoroLabeling-IPCAI2020}.

The COVID-19 lung CT scans can be accessed here: \url{https://www.imagenglab.com/newsite/covid-19/}.
    
The COVID-19 real chest X-ray data can be accessed here: \url{https://www.kaggle.com/datasets/aysendegerli/qatacov19-dataset}.
    
The COVID-19 3D lesion segmentation pre-trained network module and associated CT scans can be accessed here upon third-party restriction: \url{https://github.com/HiLab-git/COPLE-Net}.

The custom collected data for robotic surgical tool detection will be made available upon publication at \url{https://github.com/arcadelab/}.

\section*{Acknowledgements}
We gratefully acknowledge financial support from NIH NIBIB Trailblazer R21 EB028505, NIH R01 EB023939, NIH R01 EB016703 and Johns Hopkins University internal funds. 

\section*{Author contributions}
M.U. conceived of the overall idea and study design. C.G., B.D.K., and Y.H. performed data annotation, data processing, experiment setup and executed the experiments. R.G. collected the real dataset with annotations. C.G. lead result analysis and manuscript writing. All authors contributed to writing the manuscript. They also provided critical feedback, and helped shape the research and analysis.

\section*{Competing interests}
The authors have no conflict of interest to declare.

\bibliography{ref.bib}

\end{multicols}
\end{document}